\documentclass[12pt]{article}
\usepackage{epsfig,float,latexsym}

\date{}
\begin{document}

\newcommand{\beq}{\begin{equation}}
\newcommand{\eeq}{\end{equation}}
\newcommand{\nn}{\nonumber}
\newcommand{\bea}{\begin{eqnarray}}
\newcommand{\eea}{\end{eqnarray}}

\title{Computational Modelling with Modellus: An Enhancement Vector for the General University Physics Course}

\author{$\mbox{Rui Gomes Neves}^1$\footnote{For correspondence: rgn@fct.unl.pt.},$\mbox{ Jorge Carvalho Silva}^2$ \&$\mbox{ V\'{\i}tor Duarte Teodoro}^1$\\
{\small \it $\mbox{}^1$ Unidade de Investiga\c {c}\~ao Educa\c {c}\~ao e Desenvolvimento (UIED)}\\
{\small \it Departamento de Ci\^encias Sociais Aplicadas, Faculdade de Ci\^encias e Tecnologia}\\
{\small \it Universidade Nova de Lisboa}\\
{\small \it Monte da Caparica, 2829-516 Caparica, Portugal}\\
{\small \it $\mbox{}^2$ Centro de F\'{\i}sica e Investiga\c {c}\~ao Tecnol\'ogica (CEFITEC)}\\
{\small \it Departamento de F\'{\i}sica, Faculdade de Ci\^encias e Tecnologia}\\
{\small \it Universidade Nova de Lisboa}\\
{\small \it Monte da Caparica, 2829-516 Caparica, Portugal}}

\maketitle

\vspace{-0.25cm}
\begin{abstract}
In this paper we present a step forward to improve general physics as an educational experience: the implementation of a new course component composed by innovative workshop activities based on computational modelling in the general physics course taken by first year biomedical engineering students at the Faculty of Sciences and Technology of the New Lisbon University. The activities were created as interactive modelling experiments with Modellus, a computer software tool designed to construct and explore mathematical models based on functions, iterations and differential equations. Special emphasis was given to cognitive conflicts in the understanding of physical concepts, to the manipulation of multiple representations of mathematical models and to the interplay between analytical and numerical solutions of physical problems. In this work we describe these computational modelling activities and their educational aims. We also discuss their effective impact on the students learning of key physical and mathematical concepts of the course.
\end{abstract}

\section{Introduction}

General physics is an extremely difficult subject for university students taking it as a compulsory part of their first year plan of studies. Due to a lack of understanding of fundamental concepts in physics and mathematics \cite{Hestenes87,McDermott91}, the number of students that fail on the course examinations is usually very high. For example, in the Faculty of Sciences and Technology of the New Lisbon University, on average only less than 30 per cent of the students are able to take the course on the first time. What is worse is that many of those students that do actually succeed also reveal several weaknesses in their understanding of elementary physics. There are multiple reasons for this serious problem and to discuss them is beyond our present scope.

Although not exclusively, it is clear that the solution for this problem requires changes in the way physics is taught. Results from physics education research have shown that in many occasions the process of learning is effectively enhanced when students are involved in the learning activities as scientists are involved in research \cite{Mazur97}
-\cite{Beichneretal99}.

Indeed, scientific research in physics is a dynamical and exploratory process of creation, testing and improvement of mathematical models that describe observable physical phenomena. This process is an interactive blend of individual and group reflexions based on a continuously evolving set of analytical, computational and experimental techniques. It is from this cognitive frame of action that an inspiring understanding of the laws of the physical universe emerges. Hence, it should not be a surprise that physics may turn out to be more successfully taught in interactive and exploratory environments where students are helped by teachers to work as teams of scientists do. In this kind of class environment knowledge performance is better promoted and common sense beliefs as well as incorrect scientific notions can be more effectively fought.

Another important aspect of these interactive, research inspired learning environments is the possibility to give a central role to computer software tools and computational modelling. This sets the learning process in phase not only with modern scientific research where computation is as important as theory and experiment, but also with the rapid parallel development of technology.

Modelling physics in computer learning environments started with an emphasis on programming languages. Using, for example, professional languages such as Fortran \cite{Bork67}, Pascal \cite{RedishWilson93} and Python \cite{ChabaySherwood08} as well as educational languages like Logo \cite{Hurley85}, this approach requires students to develop a working knowledge of programming. The same happens when using scientific computation software such as Mathematica and Matlab. To avoid overloading students with programming notions and syntax, computer modelling systems such as Dynamic Modelling System \cite{Ogborn85}, Stella \cite{HPS97},  Easy Java Simulations \cite{ChristianEsquembre07} and Modellus \cite{Teodoro02} were developed to focus the learning activities on the understanding of the concepts of physics and mathematics. Of these, Modellus is specially interesting for allowing a model to be conceived almost as it is on a piece of paper, for having the possibility of creating animations with interactive objects that have mathematical properties expressed in the model and for permitting the analysis of experimental data in the form of images.

In this work we have followed recent results of science education research (see, e.g., \cite{Handelsmanetal05,Slootenetal06} to create and implement a new course component based on computational modelling with Modellus, in the general physics course taken by first year biomedical engineering students at the Faculty of Sciences and Technology of the New Lisbon University. In the following sections we describe how this computational modelling course was organized, what was the pedagogical methodology implemented and the educational aims of the computational modelling activities offered to the students. We also discuss their effective impact on the students learning of key physical and mathematical concepts of the general physics course.

\section{Course organization and methodology}

The general physics course for biomedical engineering involved a total of 114 students among which 50 were taking the course for the first time. The course was divided into lectures built around a set of key experiments where the general physics topics were first introduced, standard physics laboratories and the new computational modelling classes based on interactive computer workshop activities. The students taking the course for the first time had to attend all three course components while the others, already repeating the subject, were dispensed from the computational classes.

In the computational modelling classes, the students were organized in groups of two, one group for each computer in the classroom. During each class, the student teams worked on a workshop activity conceived to be an interactive and exploratory learning experience structured around a small number of problems about challenging but easily observed physical phenomena. The teams were instructed to analyse and discuss the problems on their own using the physical, mathematical and computational modelling guidelines provided by the lectures and the workshop protocols. To ensure adequate working rhythm with appropriate conceptual, analytical and computational understanding, the students were continuously accompanied and helped during the exploration of the activities. Whenever it was felt necessary, global class discussions were conducted to keep the pace and to clarify any doubts on concepts, reasoning and calculations.  

All workshop activities were created as modelling experiments based on Modellus, a computer software tool designed to introduce students and teachers to scientific computation through the construction and exploration of mathematical models based on functions, iterations and differential equations. Modellus was chosen as basis software because it is a tool that allows modelling experiments involving multiple representations (tables, graphics, and animations) almost without the need to introduce new symbols or syntax. Indeed, with Modellus it is possible to create and explore models in the computer following almost the same notation and reasoning already familiar when attempting to solve physical and mathematical problems on paper. Most importantly, Modellus allows the creation of animations with interactive objects which have their mathematical properties defined in the model and to complement them with multiple tables and graphs.

Each workshop activity consisted of a set of tasks in mechanics, presented in PDF documents, with embedded video support to help students both in class and/or at home in a collaborative online context centred on the Moodle online learning environment. The course was globally conceived to give special emphasis to cognitive conflicts in the understanding of the relevant physical concepts, to the manipulation of multiple representations of mathematical models and to the interplay between analytical and numerical approaches to the solution of physical problems.

\section{Computational modelling activities with \newline Modellus}

The program of the general physics course offered to the biomedical engineering students was based on Young and Freedman \cite{YF04}, complemented with applications of physics to biology and medicine. Following its structure the computational modelling component covered eight basic themes in mechanics \cite{Teodoro06}: 1) Vectors; 2) Motion and parametric equations; 3) Motion seen in moving frames; 4) Newton's equations: analytic and numerical solutions; 5) Circular motion and oscillations; 6) From free fall, to parachute fall and bungee-jumping; 7) Systems of particles, linear momentum and collisions and 8) Rigid bodies and rotations. During all classes, the students groups would read the PDF protocols and would follow the instructions of the embedded videos as well as our personal guidance to carry out all the modelling activities. In this section we discuss a selected set of themes from the first part of the course starting with vectors, a subject covered in the lectures just after the topics on measurable quantities and units.

\subsection{Vectors} 

\begin{figure}[H]
   \center{\psfig{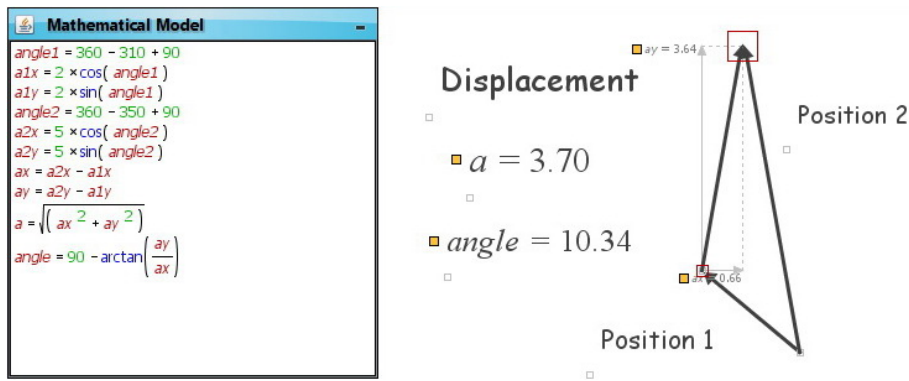}}
   \vspace{-0.2cm}
   \caption{The plane model with the animation displaying the solution. The plane has travelled 3.7 km approximately in the direction 010.}
   \label{fig1:plane}
   \end{figure}

A vector is an abstract mathematical object defined by a magnitude and a direction. In physics it is used to describe many important quantities, for example, the velocity or the force acting on a particle. With Modellus students were able to create vectors in the workspace and directly interact with them to visualise and reify many of its abstract properties. Indeed, when a vector is created its scalar and vector components are immediately visible on the screen. By simply using the computer mouse to drag the tip of the vector students were able to change its magnitude or its direction, and explore the effect on the scalar and vector components. Furthermore, introducing the vector coordinates as parameters and using Modellus predefined elementary functions, students were also taught to construct mathematical models to define the magnitude and the direction of any vector, the sum and subtraction of vectors as well as the multiplication of a vector by a scalar. In the end the students used Modellus to solve and animate physical problems involving the sum and subtraction of position vectors. An example requiring the definition of vector directions and the subtraction of two position vectors to find a displacement vector was that of a plane detected by radar (see figure~\ref{fig1:plane}): a plane is detected 2 km away in the direction 310 (using the navigation convention where the angle varies between 0 and 360 degrees clockwise starting from the North). After some time, the plane is 5 km away in the direction 350. What is the distance travelled by the plane between the two detection points? Where is the plane headed?

During class, and while attempting to solve these problems involving vectors, students revealed particular difficulties in understanding how to correctly define the direction angle of a vector in the navigation convention. At first, many students were not aware that the trigonometric functions are defined with different conventions and were frequently unable to make the vectors point in the right direction. To be able to correct the models and at the same time visualise the effect of the change in the animation, was for the students an essential advantage of the modelling process with Modellus in helping them to solve this difficulty.       

\subsection{Motion and parametric equations} 

The vector modelling activities showed students that vectors are mathematical objects used to represent physical quantities that require both a magnitude and a direction to be completely defined. An example is the velocity, a fundamental vector quantity which measures the instantaneous rate of change of the position with time. In a rectilinear and uniform motion, the velocity is constant and the position vector changes linearly with time. This type of motion can be modelled with Modellus if the coordinates of the position vector are associated with the corresponding parametric functions. In this set of activities students explored interactively several rectilinear and uniform motions on the plane using mathematical models with parametric equations, graphs of the coordinates as functions of time and particle animations representing the motion trajectory. During class, some students were not completely at ease in making the distinction between the trajectory and the graphs of the coordinates as functions of time. After completing the proposed tasks these and the other students recognized that, for allowing the possibility to visualise at the same time trajectories and different coordinate graphs, modelling with Modellus was indeed more effective in helping them to manipulate, distinguish and correctly interpret these different representations of the model. After exploring these models, students also learn how to use conditions in Modellus to define branching functions and then reproduce the corresponding motion.

\begin{figure}[H]
   \center{\psfig{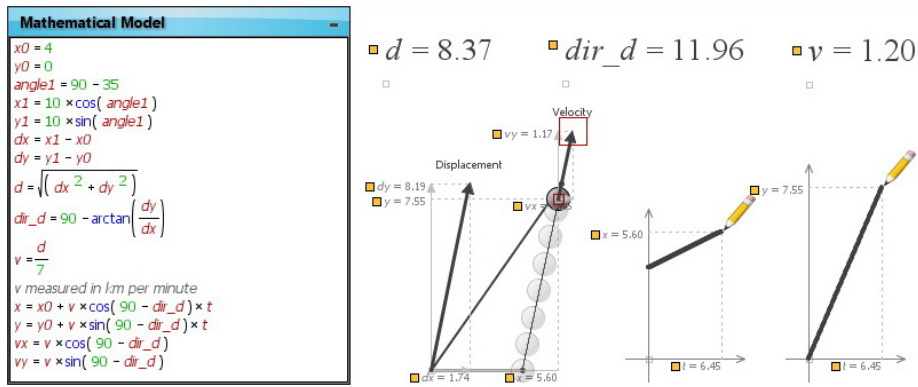}}
   \vspace{-0.2cm}
   \caption{Modelling the motion of a car. The car travelled 8.37 km approximately in the direction 012 at a speed equal to 72 km/h.}
   \label{fig2:car}
   \end{figure}

At the end of the classes covering this theme, students showed ability to complete activities requiring knowledge on how to characterise displacement vectors and velocities by their magnitude, direction and Cartesian coordinates as well as knowledge about the parametric equations of motion. For example, all the student groups were able to construct models to solve and animate the following problem about the motion of a car (see figure~\ref{fig2:car}): a car is detected 4 km away when is moving eastwards. Seven minutes later the car is found 10 km away in the direction 035. What is the distance travelled by the car between the two detection points? Where does the displacement vector points? Assuming that the motion is uniform and rectilinear, determine the velocity of the car.

\subsection{Motion seen in moving frames} 

\begin{figure}[H]
   \center{\psfig{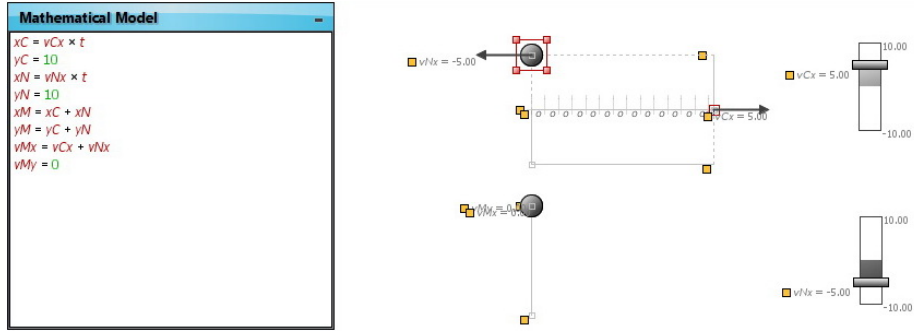}}
   \vspace{-0.2cm}
   \caption{Modelling relative motion with Modellus.}
   \label{fig3:frames}
   \end{figure}

Relative motion was the subject of the third theme of the course. During these activities students had the chance to use Modellus to experience several cognitive conflicts and realise that observers in moving reference frames can have very different views of the motion. For example, students modelled and constructed the animation of the motion of a swimmer in a river with a downstream current equal to 5 m per minute (see figure~\ref{fig3:frames}).

When the swimmer tried to move up stream with the same speed as the downstream current it would not move at all relative to an observer on the river bank. However, for an observer on a boat dragged by the current, the swimmer would move up stream with a speed of 5 m per minute. When observing this animation for the first time, students were generally startled for a moment. After some thought and careful analysis of the mathematical model associated with the animation, they were all able to understand that the two points of view are related by the Galilean velocity transformation.

\begin{figure}[H]
   \center{\psfig{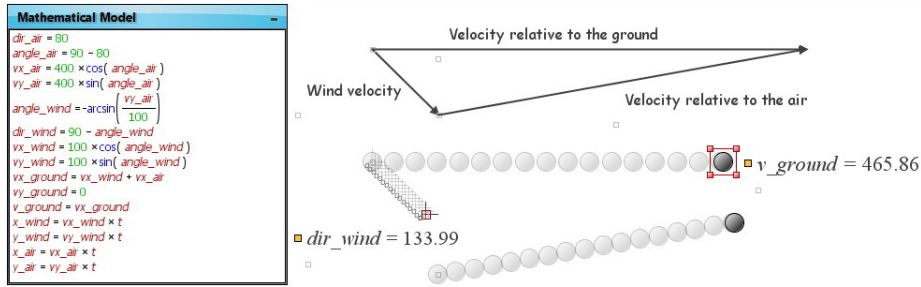}}
   \vspace{-0.2cm}
   \caption{Modelling the motion of a plane against the wind. The wind blows approximately in the direction 134 and the plane ground velocity is 465.86 km/h.}
   \label{fig4:wind}
   \end{figure}

At the end of this theme students groups were able to successfully explore with Modellus the similar problems of a boat crossing a river with current and of the motion of a plane against the wind. An example was the following: a plane flies from west to east. The pilot chooses a velocity of 400 km/h pointing in the direction 080. If the wind blows at 100 km/h, determine the wind direction and the plane ground speed (see figure~\ref{fig4:wind}). In this context, the interactive process of modelling with Modellus was of special relevance in helping students to realise that many different, everyday life physical situations can be explained using the same simple mathematical model.

\subsection{Newton's equations: analytic and numerical solutions} 

\begin{figure}[H]
   \center{\psfig{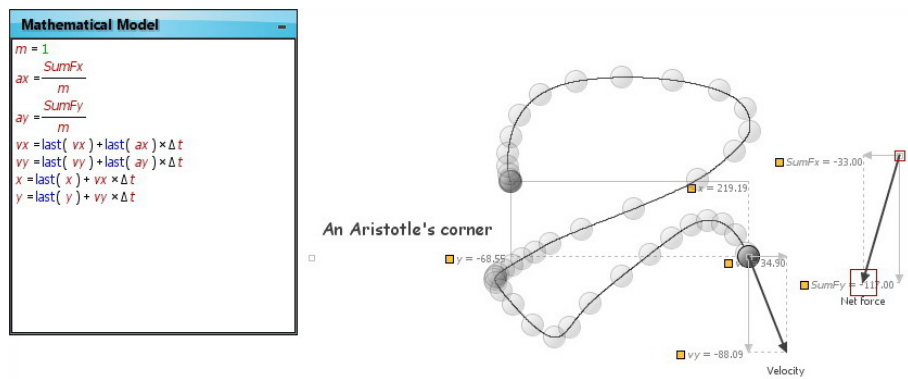}}
   \vspace{-0.2cm}
   \caption{Interactive Euler-Cromer model to make an Aristotle's corner.}
   \label{fig5:corner}
   \end{figure}

What must happen for the velocity to change during motion? This was the starting question for the fourth theme of the course. If the velocity is changing during motion there must be an acceleration vector and at least one applied force. The acceleration is the vector that measures the instantaneous rate of change of the velocity with time. According to Newton's second law of motion, this vector is obtained dividing the sum of all the forces that act on the particle by the mass of the particle. If there are no net forces then there is no acceleration and the velocity is constant. This is the statement of Newton's first law of motion or law of inertia. To explore these laws, students began with an activity where the objective was to change the velocity of a particle in the perpendicular to make an Aristotle's corner (see figure~\ref{fig5:corner}) \cite{diSessa82}. 
In the Aristotle's corner model, Newton's equations of motion are written in the form of Euler-Cromer iterations. The students were thus introduced to a simple numerical method to solve the equations of motion and determine the velocity and position of the particle knowing its mass and the net applied force. The model animation is constructed with three objects: the particle, a vector representing the velocity attached to the particle and a vector representing the net force. Because the coordinates of the net force vector are independent variables and the model is iterative, students were able to manipulate this vector at will and in real time control realistically the motion of the particle. Soon it became clear that to do an "Aristotle's corner" it was first necessary to break and stop, and only then accelerate in the perpendicular. They also learn that the choice of a small time step was important for a numerical method to work. During this modelling activity with Modellus, students were helped to resolve still another cognitive conflict: to break is not that different from accelerating, it is just to accelerate in the direction opposite to the direction of the velocity. The interactive learning process of modelling this problem in Modellus, also lead students to learn that the choice of a small time step is an important one to obtain a good simulation of the motion and that this is the same as determining a good numerical solution of the equations of motion.

\begin{figure}[H]
   \center{\psfig{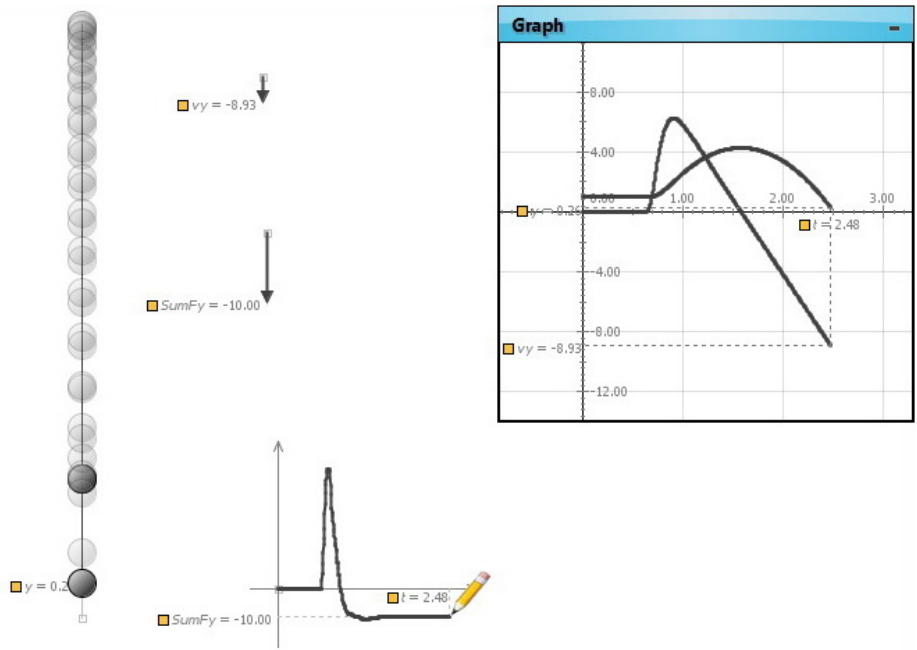}}
   \vspace{-0.2cm}
   \caption{Throwing a ball with Modellus and iterative Newton's equations.}
   \label{fig6:ball}
   \end{figure}

The same numerical model was then used to explore, for example, the throwing of a ball into the air (see figure~\ref{fig6:ball}). In this activity, students were taught to interact with the net force vector applied to the ball and to simulate the throw as well as the following motion under the earth's gravitational pull.

Again the choice of the appropriate time step was an important source of difficulties. This was resolved using the possibility of changing the mathematical model and immediately observe the effect of the correction on the animation. With the position and velocity time graphs, students were able to determine how long it takes for the ball to reach the highest point of the trajectory, what is the height of that point and when is the ball three metres up in the air. The students were also able to draw on paper the vector diagrams representing the forces acting on the ball, the velocity and the acceleration during the whole motion. This task was more easily completed when these vectors were created as animation objects. Another important learning difficulty students were helped to address during this modelling activity was the need to choose appropriate scales for the animation objects and graphs. Finally, by observing the net force graph as a function of time, they were able to estimate the duration of the throw.

\begin{figure}[H]
   \center{\psfig{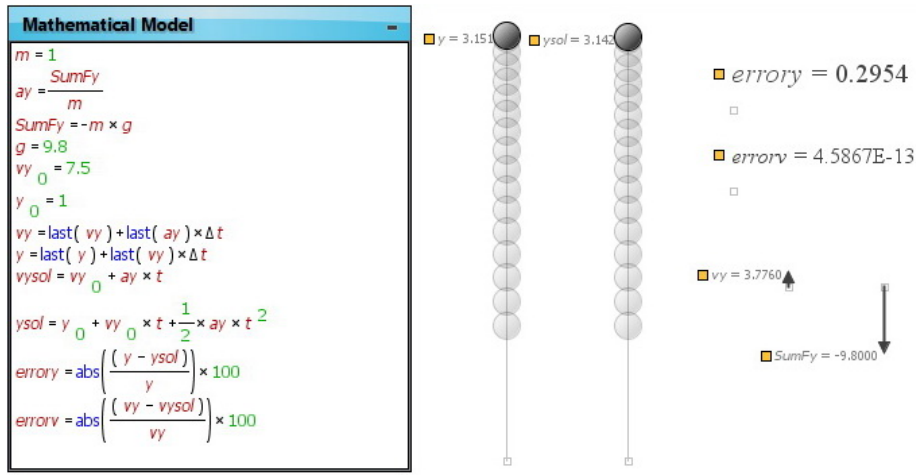}}
   \vspace{-0.2cm}
   \caption{Comparing analytic and numerical solutions to the ball throw.}
   \label{fig7:ball2}
   \end{figure}

The next activity in this theme was to compare the analytic solution for the motion just after the throw with the corresponding numerical solution obtained using the Euler method and the Euler-Cromer method (see figure~\ref{fig7:ball2}). For the same sum of applied forces and the same initial conditions students were lead to find that the analytic solution is different from the numerical solutions. They were able to understand that there is always an error associated with the iterative approximations used and learn how to quantify this error for both methods. The activity showed students that the error is only present for the position and not for the velocity, since this quantity changes linearly with time. To finish, students used Modellus to model and play a basketball game. They were also incentivised to construct the analytic solution of this projectile motion and to discuss why in such a model only the time interval after the ball leaves the hand of the player is considered.

\section{Conclusions}

In this paper we have presented the computational component of an introductory physics course based on innovative workshop activities with Modellus. After describing the course organization and methodology we have discussed the computational modelling activities to show how they were conceived to generate cognitive conflicts relative to important physical concepts, to promote the manipulation and correct interpretation of multiple representations and to analyse the interplay between analytic and numerical solutions of physical problems. We have concluded that during class the computational modelling activities with Modellus were successful in identifying and resolving several student difficulties in key physical and mathematical concepts of the course. Of fundamental importance to achieve this was the possibility to have a real time visible correspondence between the animations with interactive objects and the object's mathematical properties defined in the model, and also the possibility of manipulating simultaneously several different representations.
The implementation of this set of computational modelling activities with Modellus was thus successful. This was indeed reflected in the student answers to a questionnaire given at the end of the course \cite{Nevesetal08}. Globally, students reacted positively to the workshop activities with Modellus, considering them to be important in the context of the biomedical engineering course. Students showed a clear preference to work in teams in an interactive and exploratory learning environment, with proper guidance and support from the professors. The computational activities with Modellus presented in PDF format with embedded video guidance were also considered to be interesting and well designed. In this work Modellus was also successfully tested as a software tool that allows students to work as authors of mathematical physics models and simulations, not as simple browsers of computer simulations. Models can be presented as differential equations solved by simple numerical methods and students can appreciate the differences between numerical solutions and analytical solutions. For the students Modellus was indeed seen as helpful in the learning process of mathematical and physical models.
\vspace{0.25cm}

\leftline{\large \bf Acknowledgements}
\vspace{0.25cm}

Work supported by Unidade de Investiga\c {c}\~ao Educa\c {c}\~ao e Desenvolvimento (UIED) and Funda\c {c}\~ao para a Ci\^encia e a Tecnologia (FCT), Programa Compromisso com a Ci\^encia, Ci\^encia 2007.

\end{document}